\documentclass[onecolumn,pra,aps,nofootinbib,superscriptaddress]{revtex4-2}
\usepackage{amsfonts}
\usepackage{amssymb}
\usepackage{mathrsfs}
\usepackage{amsthm}
\usepackage{graphicx}
\usepackage{amsmath}
\usepackage{color}
\usepackage{tikz}
\usepackage{bbm}

\newtheorem{Thm}{Theorem}
\newtheorem{Prop}{Proposition}
\newtheorem{Cor}{Corollary}

\newtheorem{Lem}{Lemma}

\newtheorem{Conj}{Conjecture}

\newcommand{\Tr}[0]{\mathrm{Tr}}

\newcommand{\bei}{\begin{itemize}}
\newcommand{\eei}{\end{itemize}}

\def\<{\langle}
\def\>{\rangle}
\newcommand{{\Cn}}{{\mathbb{C}^4}}
\newcommand{{\CN}}{{\mathbb{C}^{2n}}}
\newcommand{{\BC}}{{\mathcal{B}(\mathbb{C}^n)}}
\newcommand{{\BBC}}{{\mathcal{B}(\mathbb{C}^{2n})}}

\begin{document}

\title{ A class of optimal positive maps in ${M}_n$}

\author{Anindita Bera, Gniewomir Sarbicki and Dariusz  Chru\'sci\'nski\\
\emph{Institute of Physics, Faculty of Physics, Astronomy and Informatics \\  Nicolaus Copernicus University, Grudzi\c{a}dzka 5/7, 87--100 Toru\'n, Poland}}




\begin{abstract} 
It is proven that a certain class of positive maps in the matrix algebra $M_n$ consists of optimal maps, i.e. maps from which one cannot subtract any completely positive map without loosing positivity. This class provides a generalization of a seminal Choi positive map in $M_3$.   
\end{abstract}

\maketitle

\section{Introduction}

Let $M_n$ denote a matrix algebra of $n \times n$ matrices over the complex field $\mathbb{C}$. For $X \in M_n$,  denote by $X^\dagger$ the Hermitian conjugation and by $\overline{z}$ a complex conjugation of $z \in \mathbb{C}$. For $x,y \in \mathbb{C}^n$, we denote by $\<x,y\>$ the canonical scalar product in $\mathbb{C}^n$, i.e. $\< x,y\> = x^\dagger y $.

A linear map $\Phi : M_n \to M_m$ is called positive if $\Phi(X) \geq 0$ for $X \geq 0$ \cite{Stormer1,Stormer,Bhatia,Paulsen,Evans,Tomiyama1,Tomiyama2,Tomiyama3}. 
Equivalently, $\Phi$ is positive if $\Phi(xx^\dagger) \geq 0$ for any $x \in \mathbb{C}^n$. Positive maps from $M_n$ to $M_m$ form a convex cone $\mathcal{P}_{n,m}$ and the structure of $\mathcal{P}_{n,m}$ in spite of the considerable effort is still rather poorly
understood (for some recent works see \cite{MM,OSID,CMP,KOREA,Ha-Kye,TOPICAL,Marciniak,Majewski,ani22}). Positive maps play an important role both in physics and mathematics providing generalization of $*$-homomorphisms, Jordan homomorphisms and conditional expectations. 
Moreover, it provides a powerful tool for characterizing quantum entanglement \cite{HHHH} and hence plays a key role in various aspects of quantum information theory \cite{QIT}. The notion of a positive map can be refined as follows \cite{Stormer,Paulsen,Bhatia}: $\Phi \in \mathcal{P}_{n,m}$ is $k$-positive if the extended map
\begin{equation}
  {\rm id} \otimes \Phi : M_k \otimes M_n \to M_k \otimes M_m ,
\end{equation}
is positive (cf. the recent paper \cite{Collins}). Finally, $\Phi$ is completely positive (CP) if it is $k$-positive for all $k$. Actually, ${\rm min}\{n,m\}$-positivity already guarantees CP.  A linear positive map $\Phi : M_n \to M_m$ is called decomposable if
\begin{equation}
  \Phi = \Phi_1 + \Phi_2 \circ \mathrm{T} ,
\end{equation}
where $\Phi_1$ and $\Phi_2$ are completely positive, and $\mathrm{T}$ denotes transposition in $M_n$.  It was proved by Woronowicz \cite{Woronowicz} that cones $\mathcal{P}_{2,2}$, $\mathcal{P}_{2,3}$ and $\mathcal{P}_{3,2}$ consist of decomposable maps only.
A first example of a non-decomposable map in $\mathcal{P}_{3,3}$ was provided by Choi \cite{Choi0,Choi1,Choi2,Choi3}
\begin{equation}
\label{Choi}
  \Phi(X) = \begin{bmatrix}
   x_{00} + x_{11} &-x_{01} & -x_{02} \\ -x_{10} & x_{11}+x_{22} & - x_{12} \\ -x_{20} & -x_{21} & x_{22}+x_{00} 
\end{bmatrix},
\end{equation}
with $X =(x_{ij}) \in M_3$. Interestingly Choi map turns out to be extremal in $\mathcal{P}_{3,3}$ (cf. also \cite{Ha-extr}). In this paper, we analyze another property of positive maps called optimality. Recall that  $\mathcal{P}_{n,m}$ contains a convex cone $\mathcal{CP}_{n,m}$ of completely positive maps. Now, a map $\Phi \in \mathcal{P}_{n,m}$
is called optimal if for any map $\Psi \in \mathcal{CP}_{n,m}$ the map $\Phi - \Psi$ is no longer positive, i.e. does not belong to $\mathcal{P}_{n,m}$. This notion was introduced in mathematical physics in \cite{Lew} in connection to quantum entanglement. It is clear that optimality is less restrictive than extremality. Any extremal map is optimal but the converse needs not be true  \cite{KOREA,TOPICAL}. A simple example is provided by so-called reduction map in $\mathcal{P}_{n,n}$
\begin{equation}   \label{RED}
    R_n(X) = \mathbf{I}_n {\rm Tr}\,X - X,
\end{equation}
where $\mathbf{I}_n$ stands for the identity matrix in $M_n$. It is well known \cite{KOREA,TOPICAL} that $R_n$ is optimal for all $n\geq 2$ but it is extremal for $n=2$ only.  Authors of \cite{Lew} provided the following sufficient condition for optimality.

\begin{Thm}\label{TH1} Let $\Phi \in \mathcal{P}_{n,m}$. Consider a family of product vectors $x_i \otimes y_i \in \mathbb{C}^n \otimes \mathbb{C}^m$ such that
\begin{equation}\label{xy=0}
  \< y_i, \Phi( \overline{x}_i \overline{x}_i^\dagger) y_i \> = 0 .
\end{equation}
If  $\{x_i \otimes y_i\}$ span $\mathbb{C}^n \otimes \mathbb{C}^m$, then $\Phi$ is optimal. 
\end{Thm} 
Maps possessing a full spanning set $\{x_i \otimes y_i\}$ satisfying (\ref{xy=0}) are said to have {\it spanning property} \cite{Lew}. Spanning property is therefore sufficient for optimality. However, there exist optimal maps without spanning property \cite{S71,S72,Remik,PRA-2022}. Again the Choi map serves as an example since there exists only 7 linearly independent product vectors $\{x_i \otimes y_i\}$ and the full spanning requires in this case 9 vectors \cite{S71,S72}.

In this paper, we analyze optimality of a large class of positive maps in $\mathcal{P}_{n,n}$ which provides a generalization of the Choi map from $\mathcal{P}_{3,3}$.  It is proven that maps in a particular subclass is optimal (cf. Theorem \ref{TH-M}). Interestingly, these maps do not possess a spanning property (cf. Theorem \ref{TH-S}). It is conjectured how to optimize maps outside this special subclass.

\section{Class of maps}

In this section, we study a class of positive maps in $\mathcal{P}_{n,n}$ being a generalization of the Choi map (\ref{Choi}). Let 
 $\varepsilon: M_n \to M_n$ be the canonical projection of $M_n$ to the diagonal part
\begin{equation}
\varepsilon(X)=\sum_{i=0}^{n-1} \Tr[X e_{ii}] e_{ii},
\end{equation}
where $e_{ij}$ denotes the matrix units in $M_n$. Let $\{e_0,e_1,\ldots,e_{n-1}\}$ be the canonical orthonormal basis in $\mathbb{C}^n$ and let $\sigma$ be a  permutation  defined by
\begin{equation}
\sigma e_i=e_{i+1},  \ \ \ \ \ (\mbox{mod}~n),
\end{equation}
for $i=0,1,\ldots, n-1$. The following  maps $\tau_{n,k} : M_n \to M_n$
\begin{equation}  \label{!}
\tau_{n,k}(X)=(n-k) \varepsilon(X)+\sum_{i=1}^{k} \varepsilon\big(\sigma^i X \sigma^{\dagger i}\big)-X, ~~~X\in M_n,
\end{equation}
for $k=0,1,\ldots,n-1$, were proved to be positive \cite{Ando,RIMS,yamagami}. 
From now on, the summation in the indices of matrices are considered to be mod $n$.
It is easy to see that $\tau_{n,0}$ is completely positive and $\tau_{n,n-1}$ is nothing but the reduction map (\ref{RED}) 
which is completely copositive (i.e. $R_n$ composed with transposition is completely positive).
It has been shown \cite{RIMS} that the map $\tau_{n,k}$ is  atomic and non-decomposable for $n\geq 3$ and $k=1,2,\ldots, n-2$. Recall, that a map is said to be an atomic map if it cannot be decomposed into a sum of a 2-positive and 2-copositive maps.

Let us write the map $\tau_{n,k}(X)$ explicitly in the following matrix form
\begin{equation}\label{F}
  [\tau_{n,k}(X)]_{ii} = (n-k-1) x_{ii} + x_{i+1,i+1} + \ldots + x_{i+k,i+k} \ , \ \ \ \  [\tau_{n,k}(X)]_{ij} = - x_{ij} , \ \ (i\neq j) .
\end{equation}
Formula (\ref{F}) reduces to (\ref{Choi}) for $n=3$ and $k=1$. Interestingly, the family of maps (\ref{!}) satisfies the following covariance property
\begin{equation}
  U \tau_{n,k}(X) U^\dagger = \tau_{n,k}(UXU^\dagger) ,
\end{equation}
where $U \in M_n$ is an arbitrary diagonal unitary matrix. Hence $\tau_{n,k}$ is covariant w.r.t. maximal commutative subgroup of the {special} unitary group $SU(n)$. Covariant maps were recently analyzed in \cite{Marek,Nechita1,Nechita2}. 

\section{Spanning property for the maps $\tau_{n,k}$}

Denote by  $\Sigma_n$ a linear subspace of $\mathbb{C}^n \otimes \mathbb{C}^n$ spanned by vectors $x \otimes \overline{x}$, where  $x = (e^{it_0},\ldots,e^{it_{n-1}})^T$ with real phases $t_k$.
\begin{Lem}
Any vector $y \in \mathbb{C}^n \otimes \mathbb{C}^n$ orthogonal to $\Sigma_n$ has the following form
\begin{equation}\label{definey}
  y = \sum_{k=0}^{n-1} y_k e_k \otimes e_k ,
\end{equation}
with $\sum_k y_k = 0$.  
\end{Lem}
It is clear that the subspace $\Sigma^\perp_n$ of such $y$'s is $(n-1)$-dimensional and hence ${\rm dim}~\Sigma_n = n^2-n+1$. 

\begin{Lem} If $x = (e^{it_0},\ldots,e^{it_{n-1}})^T$ with real $t_k$, then 
\begin{equation}\label{xxxx}
  \overline{x}^\dagger \tau_{n,k}(\overline{x}\overline{x}^\dagger) \overline{x} = 0.
\end{equation}
\end{Lem}
One easily checks (\ref{xxxx}) by direct calculation. It is, therefore, clear that one has at least $n^2-n+1$ vectors $x \otimes y$ satisfying (\ref{xy=0}). Now, we show that actually there are no more linearly independent vectors with this property apart from the reduction map \cite{Justyna}. Particularly, in the below theorem, we now show that if $k < n-1$, then the map $\tau_{n,k}$ does not have a spanning property. 


\begin{Thm}  \label{TH-S} Let $k < n-1$. If 
\begin{equation}\label{span1}
  y^\dagger \tau_{n,k}(\overline{x}\overline{x}^\dagger) y = 0 , 
\end{equation}
then $x \otimes y \in \Sigma_n$.
\end{Thm}
Proof: note that if $\{x \otimes y \}$ satisfies (\ref{span1}), then  $y \in \ker\{\tau_{n,k}(\overline{x} \overline{x}^\dagger)\} \subset \mathbb{C}^n \otimes \mathbb{C}^n$.
The kernel is non-trivial if and only if
 \begin{equation}
 \label{det2}
    \det[{\tau_{n,k}(\overline{x} \overline{x}^\dagger)}]=0.  
 \end{equation}
The  matrix $\tau_{n,k}(\overline{x} \overline{x}^\dagger)$  has the following form
\begin{equation}
    \tau_{n,k}(\overline{x} \overline{x}^\dagger) = A - B ,
\end{equation}
where
\begin{eqnarray}
\label{formA}
A=  \mathrm{Diag}[D_0,D_1,\ldots,D_{n-1}] , 
\label{map2}
\end{eqnarray}
with
\begin{equation}
    D_i = (n-k) X_i+X_{i+1}+\ldots+X_{i+k}, \ \ \ \  X_i:=|x_i|^2 ,
\end{equation}
and $B= \overline{x} \overline{x}^\dagger$.
Let us represent the matrices $A$ and $B$ via the corresponding columns
$$ A=[A_0~|~A_1~|\ldots|~A_{n-1}] \ , \ \ \ B=[B_0~|~B_1~|\ldots |~B_{n-1}]. $$
Note that the determinant  is a multilinear function in each column, hence one obtains $2^n$  summands. Simple algebra leads to 
\begin{eqnarray}
\label{det1}
    \det\big[\tau_{n,k}(\overline{x} \overline{x}^\dagger)\big] &=& \det\big[~A_0-B_0~|~A_1-B_1~|\ldots |~A_{n-1}-B_{n-1}~|~\big] \nonumber\\
    &=& \det\big[A_0|A_1|\ldots|A_{n-1}|\big]-\det\big[B_0|A_1|A_2|\ldots|A_{n-1}|\big]-
    \ldots-\det\big[A_0|A_1|\ldots|A_{n-2}|B_{n-1}|\big] \nonumber\\ 
    &=&  
  \prod_{i=0}^{n-1} D_i \, -\, \sum_{j=0}^{n-1} X_j \prod_{\substack{i=0 \\ i\neq j}}^{n-1} D_i . 
\end{eqnarray}
\textbf{Case 1 $(\det A > 0)$:}  If the determinant  $\det A = \prod_{i=0}^{n-1} D_i >0 $,  i.e. 
\begin{equation}
\label{connew1}
D_i = (n-k)X_i+X_{i+1}+\ldots+X_{i+k}>0 \ , \ \ \ \ i=0,1,\ldots,n-1 , 
\end{equation}
then formula (\ref{det1}) implies
\begin{eqnarray}
\label{newdet1}
    \det\big[\tau_{n,k}(\overline{x} \overline{x}^\dagger)\big] = \det A \left( 1 - \sum_{j=0}^{n-1} \frac{X_j}{D_j} \right) .
\end{eqnarray}
Assuming $D_i >0$ for all $i$, let us define a function
\begin{equation}
\label{eq:fx}
 f_{n,k}(X_0,\ldots,X_{n-1}) = \sum_{i=0}^{n-1}  \frac{X_i}{D_i} = 
 \sum_{i=0}^{n-1} \frac{X_i}{(n-k) X_i+X_{i+1}+\ldots+X_{i+k}} .
\end{equation}
Actually, positivity of $\tau_{n,k}$ implies \cite{yamagami} that $f_{n,k} \leq 1$. One finds that $ \det\big[\tau_{n,k}(\overline{x} \overline{x}^\dagger)\big]=0$ if and only if
\begin{equation}  
\label{f=1}
  f_{n,k}(X_0, \ldots, X_{n-1}) = 1 .
\end{equation}

 To find the  maximum of the function $f_{n,k}$ in the region $(X_0,\ldots,X_{n-1})$ where $D_i >0$, one has to analyze the property of the corresponding Hessian matrix. One finds that 
\begin{equation}
     \frac{\partial^2 f_{n,k}(X_0,\ldots,X_{n-1})}{\partial X_i \partial X_j} = - \hat{S}_{ij}.
\end{equation}
 The matrix $\hat{S}=[\hat{S}_{ij}]$ is defined as follows
\begin{equation}\label{hat-S}
	    \hat S=s'(S + S^T) - 2 S^T S,
	\end{equation}
 where the matrix $S$ is actually the matrix $A$ defined in \eqref{formA},
 and $s'>0$ is the common eigenvalue of $S$ and $S^T$ corresponding to the eigenvector $\mathbbm{1}=(1,1,\ldots,1)$.
 Then the function $f_{n,k}(X_0,\ldots,X_{n-1})$ has  a unique extremum attained at $X_0=X_1=\ldots=X_{n-1}$. This is due to the following lemma mentioned by Yamagami \cite{yamagami}:
 \begin{Lem}
\label{yama}
Let $S=[S_{ij}]$ be an invertible $n \times n$ matrix with non-negative real entries such that  $S$ and its transpose $S^T$ admit  $\mathbbm{1}=(1,1,\ldots,1)$ as an eigenvector corresponding to the common eigenvalue $s'>0$.
If the matrix $\hat{S}$ defined in (\ref{hat-S}) is positive semidefinite and its kernel is spanned by $\mathbbm{1}=(1,1,\ldots,1)$, then $\mathbbm{1}$ is a unique (up to scalar factor) point that gives a local maximum in the region $(X_0,\ldots,X_{n-1})$ where $D_i >0$, of the function $f_{n,k}$ defined in Eq.~(\ref{eq:fx}).
\end{Lem}

Due to Lemma 6 in \cite{yamagami}, the matrix $\hat{S}$ is positive semidefinite.  Therefore, in the region $(X_0,\ldots,X_{n-1})$ where $D_i >0$,  the function $f_{n,k}(X_0,\ldots,X_{n-1})=1$ if and only if
$$   X_0=X_1 = \ldots = X_{n-1} ,  $$
and hence (up to a factor) one has  
    $x=(e^{i t_0},\ldots,e^{it_{n-1}})$. 
Note, that for such vector, $\tau_{n,k}(\overline{x} \overline{x}^\dagger) = ||x||^2 \mathbf{I}_n - \overline{x} \overline{x} ^\dagger$.
%
Hence, $ [\tau_{n,k}(\overline{x} \overline{x}^\dagger)]y=0$ if and only if $y = \overline{x}$. This proves that $x \otimes y = x \otimes \overline{x} \in \Sigma_n$.  
\vspace{0.5cm}

\textbf{Case 2 ($\det A = 0$):}
Consider now the complementary region, i.e. when there exists at least one $i$ such that $D_i=0$. Suppose, for example,  $D_0=0$ and $D_j >0$ for $j>0$. It implies 
\begin{equation}
\label{connew2}
    X_0=X_1=\ldots=X_k=0 .
\end{equation}
One finds the following block diagonal structure
\begin{eqnarray}
\tau_{n,k}(\overline{x} \overline{x}^\dagger)=
\left[
\begin{array}{c|c|c}
   A' &  & \\ 
\hline   & A'' &   \\ 
\hline   &  & A'''
 \end{array}
\right]-
\left[
\begin{array}{c|c|c}
   B' &  & \\ 
\hline   & B'' &   \\ 
\hline   &  & B'''
 \end{array}
\right].
\end{eqnarray}
First $1 \times 1$ blocks $A'$ and $B'$ vanish.  The second $k \times k$ block $A''$ has the following form:

$$ A''=\mathrm{Diag}[X_{k+1},X_{k+1}+X_{k+2},\ldots, X_{k+1}+\ldots+X_{2k}] , $$
whereas the other  $k \times k$ block  $B''=0$.
The remaining $(n-k-1) \times (n-k-1)$ blocks $A'''$ and $B'''$ have the following form
\begin{eqnarray}
    A'''=\mathrm{Diag}\Big[(n-k) X_{k+1}+X_{k+2}+\ldots+X_{2k+1},\ldots, (n-k) X_{n-k}+X_{n-k+1}+\ldots+X_{n-1},\ldots, \nonumber\\
  (n-k)X_{n-2}+X_{n-1},(n-k) X_{n-1}\Big], 
 \end{eqnarray}
 and 
 \begin{eqnarray}
   B'''=\begin{bmatrix}
   X_{k+1} & \overline{x}_{k+1} x_{k+2} & \ldots & \ldots  & \overline{x}_{k+1} x_{n-1} \\ 
   \overline{x}_{k+2} x_{k+1} & X_{k+2} &  \overline{x}_{k+2} x_{k+3} &\ldots &  \overline{x}_{k+2} x_{n-1}\\ 
 \vdots  & \vdots &  \ddots & \vdots & \vdots \\ 
   \overline{x}_{n-1} x_{k+1} &  \overline{x}_{n-1} x_{k+2} & \ldots & \ldots & X_{n-1}
 \end{bmatrix}.  
\end{eqnarray}

Now, since $A'=B'=0$, one has $y=[1,0,0,\ldots,0]^T \in \ker \tau_{n,k}(\overline{x} \overline{x}^\dagger)$ and additional spanning vectors appear. Observe however, that 
$   x \otimes y \in \Sigma_n$, 
since $x \in {\rm span}\{e_{k+1},\ldots,e_{n-1}\}$ and $y = e_0$ and hence $x \otimes y \perp e_i \otimes e_i$ for all $i=0,1,\ldots,n-1$.

Now, the second block $A''-B''=A''$ is positive definite and provides no additional spanning vectors. Finally, the determinant of the third block, $A'''-B'''$
reads
 \begin{eqnarray}
  \det\big[\tau_{n,k}(\overline{x} \overline{x}^\dagger)\big] &=& 1-\frac{X_{k+1}}{(n-k) X_{k+1}+X_{k+2}+\ldots+X_{2k+1}}-\ldots \\ &-& \frac{X_{n-k}}{(n-k) X_{n-k}+X_{n-k+1}+\ldots+X_{n-1}}-\ldots-
      \frac{X_{n-2}}{(n-k) X_{n-2}+X_{n-1}}-\frac{X_{n-1}}{(n-k)X_{n-1}} . \nonumber
 \end{eqnarray}
It is bounded from below by the following quantity 
 \begin{eqnarray}   \label{no-red}
  \det\big[\tau_{n,k}(\overline{x} \overline{x}^\dagger)\big]   \geq
     1 -
     \frac{X_{k+1}}{(n-k) X_{k+1}}-\ldots-\frac{X_{n-k}}{(n-k) X_{n-k}}-\ldots- \frac{X_{n-2}}{(n-k) X_{n-2}}- \nonumber\\
     \frac{X_{n-1}}{(n-k)X_{n-1}}
     \geq
     1 - \frac{n-k-1}{n-k} >0,
 \end{eqnarray}
 where we assumed that $k< n-1$. Hence the matrix  $A'''-B'''$ is positive definite which implies the trivial kernel and no additional spanning vectors. 
 

If more diagonal elements vanish, the block $A'-B'$ will be bigger, but still the spanning vectors it provides belong to $\Sigma_n$. The blocks $A'' - B''$ and $A''' - B'''$ are again non singular and provide no spanning vectors. 
Therefore we  conclude that the total number of linearly independent  spanning vectors for the map $\tau_{n,k}$ is $n^2-n+1$ and all of them belong to $\Sigma_n$. \hfill $\Box$

It should be stressed that the very condition $k< n-1$ is essential for the proof, that is, the theorem is not true for the reduction map $R_n = \tau_{n,n-1}$ which has the spanning property \cite{Justyna}. Hence, apart $R_n$ all maps $\tau_{n,k}$ do not have the spanning property. For example, the original Choi map $\tau_{3,1}$ allows  only 7 linearly independent vectors satisfying (\ref{xy=0}) in $\mathbb{C}^3 \otimes \mathbb{C}^3$ \cite{S71,S72}.

\section{Optimality of the map $\tau_{n,k}$}

Maps $\tau_{n,k}$ do not have a spanning property which is sufficient for optimality \cite{Lew}. Interestingly, there exists a special subclass which is nevertheless optimal.   
Let us denote the greatest common divisor of $n$ and $k$ by ${\rm gcd}(n,k)$. The main result of this section consists in the  following

\begin{Thm}
\label{TH-M} 
If ${\rm gcd}(n,k)=1$, then $\tau_{n,k}$ is optimal. 
\end{Thm}

Note that if $\tau_{n,k}$ is not optimal, then one can always subtract some CP map $\Lambda$ such that $\tau_{n,k} - \Lambda$ is still positive. Since
$\overline{x}^\dagger \tau_{n,k}( \overline{x} \overline{x}^\dagger) \overline{x} = 0$
for all $x \otimes \overline{x} \in \Sigma_n$, 
then the CP map $\Lambda$ needs to satisfy:  $
\overline{x}^\dagger \Lambda( \overline{x} \overline{x}^\dagger) \overline{x} = 0$ 
for all $x \otimes \overline{x} \in \Sigma_n$.

\begin{Prop}  \label{PI} Let $\Lambda : M_n \to M_n$ be a CP map. One has

\begin{equation}\label{L=0}
  \overline{x}^\dagger \Lambda( \overline{x} \overline{x}^\dagger) \overline{x} = 0 ,
\end{equation}
for all $x \otimes \overline{x} \in \Sigma_n$ if and only if
\begin{equation}\label{lambdax}
  \Lambda(X) = \sum_{k,l=0}^{n-1} L_{kl} \,  e_{kk} X e_{ll} 
\end{equation}

together with 
$\sum_{k,l} L_{kl} = 0$,
that is,
$\Lambda(X) =L\circ X$ 
it is a Hadamard product of $X$ with a positive semi-definite matrix $[L_{kl}]$.
\end{Prop}
Proof: Consider an arbitrary CP map represented as follows \cite{Choi-75}

\begin{equation}\label{}
  \Lambda(X) = \sum_{i,j,k,l} C_{ij,kl} e_{ij} X e_{kl}^\dagger ,
\end{equation}
with positive semi-definite matrix $[C_{ij,kl}] \in M_n \otimes M_n$. Actually, $C_{ij,kl}$ is the Choi matrix of the map $\Lambda$ \cite{Choi-75}. Taking $x = (e^{it_0},\ldots,e^{it_{n-1}})^T$ with real $t_k$, one finds
\begin{equation}\label{}
  \Lambda(\overline{x}\overline{x}^\dagger) = \sum_{i,j,k,l} C_{ij,kl} e^{-i(t_j -t_l)} e_{ik}  ,
\end{equation}
and hence
\begin{equation}\label{}
  \overline{x}^\dagger \Lambda( \overline{x} \overline{x}^\dagger) \overline{x} =  \sum_{i,j,k,l} e^{i(t_i -t_j)} C_{ij,kl}  e^{-i(t_k -t_l)} .
\end{equation}
Now, in order to have $ \overline{x}^\dagger \Lambda( \overline{x} \overline{x}^\dagger) \overline{x}=0$ for all possible $t_k$, one has
\begin{equation}\label{}
  C_{ij,kl} = L_{ik} \delta_{ij} \delta_{kl}+a_{ij} \delta_{ik} \delta_{jl} (1-\delta_{ij}) ,
\end{equation}
The remaining entries of $C_{ij,kl}$ are multiplied by a non-vanishing arbitrary phase hence has  to vanish and consequently, we have the following equation
\begin{equation}\label{}
  \overline{x}^\dagger \Lambda( \overline{x} \overline{x}^\dagger) \overline{x} =  \sum_{i,k} L_{ik}+\sum_{i \neq j} a_{ij} ,
\end{equation}
and the condition (\ref{L=0}) implies $ \sum_{i,k} L_{ik}+\sum_{i \neq j} a_{ij}=0 $. 
%

On the other hand, $a_{ij}$ are non-negative as diagonal elements of $C$ and the matrix $L$, being a diagonal block of $C$ is positive semi-definite, in particular $\mathbbm{1}^\dagger L \mathbbm{1}  = \sum_{i,k} L_{ik} \ge 0$. These implies $\sum_{i,k} L_{ik} = 0$ and $\forall i \ne j, \ a_{ij}$ = 0. In this way we obtain that the map $\Lambda$ is of the form \eqref{lambdax} and $\Lambda$
is a completely positive map if and only if the matrix $L=[L_{kl}]$ is positive semi-definite \cite{Bhatia,Horn}.

 \hfill $\Box$

Proof of the Theorem \ref{TH-M}: {If the map $\tau_{n,k}$ is not optimal one can always subtract a  CP map satisfying the properties of Proposition \ref{PI} with a rank one positive matrix $C = \alpha \alpha^\dagger$ such that}
\begin{equation}\label{taualpha}
\tau_{n,k}^{(\alpha)}(X) :=  \tau_{n,k}(X)  - \alpha\alpha^\dagger \circ X \geq 0 , 
\end{equation}
for all $X \geq 0$, where $\sum_i \alpha_i = 0$. Consider now $X = uu^\dagger$, where
\begin{equation}\label{u}
  u = (1,s,s^2,\cdots,s^{n-k-1},\underbrace{0, \ \cdots \ ,0}_k)^T ,
\end{equation}
and $s \in \mathbb{R}$. One finds
\begin{equation}\label{}
\tau_{n,k}^{(\alpha)}(X) = \begin{bmatrix} A & 0 \\ 0 & B 
      \end{bmatrix} ,
\end{equation}
where $B \in M_{k}$ is evidently positive semi-definite, and $A \in M_{n-k}$ is defined as follows
\begin{equation}
A =  \sum_{i=0}^{n-k-1}\Big[ (n-k)s^{2i}+\sum_{j=i+1}^{\min\{i+k,n-k-1\}} s^{2j}\Big] e_{ii} -  P(X + \alpha\alpha^\dagger \circ X)P ,
\end{equation}
with $P:M_n \rightarrow M_{n-k}$, canonical embedding.

It can be written as follows
\begin{eqnarray}
   A =  \mathrm{Diag}[1,s,s^2,\cdots,s^{n-k-1}]\, \widetilde{A} \ 
     \mathrm{Diag}[1,s,s^2,\cdots,s^{n-k-1}],
\end{eqnarray}
where

\begin{equation}\label{}
\widetilde{A} = 
    \sum_{i=0}^{n-k-1} \Big[ (n-k)+\sum_{j=i+1}^{\min\{i+k,n-k-1\}} s^{2(j-i)}\Big] e_{ii}
    - \mathbb{J}_{n-k}     - \widetilde{\alpha} \widetilde{\alpha}^\dagger  ,
\end{equation}

$\widetilde{\alpha} = P\alpha$, and  $\mathbb{J}_m \in M_m$ is defined via $(\mathbb{J}_m)_{kl} = 1$. Clearly, $A \geq 0$ if and only if $\widetilde{A} \geq 0$.
If the map $\tau_{n,k}^{(\alpha)}$ is positive, then $\widetilde{A}$  has to be positive for all $s$, hence it is positive in the limit $s \to 0$
\begin{eqnarray}
     (n-k) \mathbb{I}_{n-k} - \mathbb{J}_{n-k} - \widetilde{\alpha} \widetilde{\alpha}^\dagger \geq 0 .
\end{eqnarray}
Note that the matrix $(n-k) \mathbb{I}_{n-k} - \mathbb{J}_{n-k}$ is positive semi-definite and
\begin{equation}\label{}
  [(n-k) \mathbb{I}_{n-k} - \mathbb{J}_{n-k}] \beta = 0 ,
\end{equation}
with $\beta =(1,\ldots,1)^T$. Hence, a necessary condition for positivity of $\widetilde{A}$ is orthogonality of $\beta$ and $\widetilde{\alpha}$, that is,
\begin{eqnarray}
   \beta^\dagger \tilde{\alpha} =  \sum_{i=0}^{n-k-1} \alpha_i =0.
\end{eqnarray}
Similar analysis may be performed for a vector $\sigma^j u$ (with $u$ defined in (\ref{u})). It implies
\begin{eqnarray}
    \sum_{i=j}^{n+j -k-1} \alpha_i =0 ,
\end{eqnarray}
for arbitrary $j=0,1,\ldots, n-1$. Hence, the vector $\alpha \in \mathbb{C}^n$ has to satisfy the following condition
\begin{equation}
\label{c1}
  \mathbf{M} \alpha = 0 ,
\end{equation}
where $\mathbf{M}$ is the following circulant matrix 
\begin{equation}
\mathbf{M} = \begin{bmatrix} m_{0} & m_{1} & \ldots & m_{n-1} \\ m_{n-1} & m_0 & \ldots & m_{n-2} \\
\vdots & \vdots & \ddots & \vdots \\
m_1 & m_2 & \ldots & m_0 \end{bmatrix} ,
\end{equation}
with 
$$  m_0 = \ldots =  m_{n-k-1}=1 \ , \ \ \ m_{n-k}=\ldots m_{n-1}= 0 . $$ 
It is, therefore, clear that  $\alpha$, for which $\tau_{n,k}^\alpha$ in \eqref{taualpha} is positive, defines an eigenvector of $\mathbf{M}$ to zero eigenvalue. The eigenvalues of the circulant matrix $\mathbf{M}$ read as follows
\begin{equation}
    \lambda_j=1+\omega^j+\omega^{2 j}+ \ldots +\omega^{(n-k-1)j},
\end{equation}
for $j=0,1, \ldots,n-1$, with $\omega=e^{2\pi i/n}$.
Now, using 
\begin{equation}
  1+\omega^j+\omega^{2 j}+ \ldots +\omega^{(n-1)j}=0 ,
\end{equation}
one obtains
\begin{eqnarray}
  \lambda_j &=&  1+\omega^j+\omega^{2 j}+ \ldots +\omega^{(n-k-1)j} = -\Big(\omega^{(n-k)j}+\omega^{(n-k+1)j}+\ldots +\omega^{(n-1)j}\Big)\nonumber\\
    &=& -\omega^{(n-1)j} \Big(1+(1/\omega)^j+ \ldots +(1/\omega)^{(k-1)j} \Big) =:  -\omega^{(n-1)j} \cdot g_j .
\end{eqnarray}
Hence $\lambda_j=0$ if and only if  $g_j=0$. Assuming that $j\ne 0$ one has that $g_j=0$ if and only if
\begin{equation}
    \Big( (1/\omega)^j-1 \Big) g_j =
    (1/\omega)^{kj}-1
    = 0.
\end{equation}
Observe that $\lambda_0 = n-k$. If $j \neq 0$, we have the following condition
\begin{equation}
\label{new1}
    k \cdot j=0 , \ \ \ \ \  ({\rm mod} ~n).
\end{equation}
The above linear congruence has exactly ${\rm gcd}(n,k)$ solutions for a number `$j$'. In particular, there are ${\rm gcd}(n,k)-1$ non-zero solutions.  It implies that, if $n$ and $k$ are relatively prime, then the matrix $\mathbf{M}$ in Eq.~(\ref{c1}) is non-singular and there are no  non-zero solutions of Eq.~(\ref{c1}). Hence,  there does not exist a vector $\alpha \neq 0$ such that the map $\tau^{(\alpha)}_{n,k}(X)  = \tau_{n,k}(X) - \alpha\alpha^\dagger \circ X$ remains positive. It proves, therefore, that the map $\tau_{n,k}$ is optimal whenever ${\rm gcd}(n,k)=1$. \hfill $\Box$

In the appendix \ref{simple}, we show the optimality for three simple examples: $\tau_{3,1}$, $\tau_{4,3}$ and $\tau_{5,3}$ .

\begin{Cor}  Note that $\mathrm{gcd}(n,1)=1$ for any $n$. Hence $\tau_{n,1}$ is always optimal. Choi map $\tau_{3,1}$ serves as an example.  
If $k=n-2$ and $n$ is odd, then $\mathrm{gcd}(n,n-2)=1$ and hence $\tau_{n,n-2}$ is optimal. In particular, it applies for the original Choi map $\tau_{3,1}$. 
Moreover, $\mathrm{gcd}(n,n-1)=1$ for any $n$ and hence $\tau_{n,n-1}$ is always optimal. Since $\tau_{n,n-1} = R_n$ (reduction map) this result recovers well known fact that $R_n$ is optimal for any $n\geq 2$. It is extremal only for $n=2$.
\end{Cor}

\section{Beyond optimality: $\mathrm{gcd}(n,k) \geq 2$} 

In general, if $d={\rm gcd}(n,k)$, then the linear congruence (\ref{new1}) has exactly $d-1$ non-zero solutions:
\begin{equation}
    j = r \cdot (n/d) \ \ \ ({\rm mod} ~n),\ \ \ r =1,2,\ldots, d-1 ,
\end{equation}
and exactly $d-1$ eigenvalues $\lambda_j$'s are equal to zero. The corresponding eigenvectors read
\begin{equation}
\label{vectors}
    v_r = \frac{1}{\sqrt{n}} \Big(1, \omega^{\frac{n}{d}r},(\omega^{\frac{n}{d}r})^2, \ldots,(\omega^{\frac{n}{d}r})^{n-1}\Big)^T, \ \ \ r=1,2,\ldots, d-1 .
\end{equation}
The subspace  spanned by $\{v_1,\ldots,v_{d-1}\}$ defines  the kernel of $\mathbf{M}$. In this case any nontrivial solution of (\ref{c1}) is a linear combination of $v_r$. Assuming normalization $\| \alpha_r \|=1$, one may expect that
\begin{equation}\label{}
  \tau_{n,k}(X) - \Big(  a_1 \alpha_1 \alpha_1^\dagger \circ X + \ldots + a_{d-1} \alpha_{d-1} \alpha_{d-1}^\dagger \circ X \Big) ,
\end{equation}
gives rise to an optimal positive map for a suitable choice of non-negative parameter $\{a_1,\ldots,a_{d-1}\}$.

The simplest scenario corresponds to $d={\rm gcd}(n,k)=2$. In this case there exists a single eigenvector $v_1$ spanning the kernel of $\mathbf{M}$
\begin{eqnarray}
    v_1 = \frac{1}{\sqrt{n}} \Big(1, \omega^{\frac{n}{d}},(\omega^{\frac{n}{d}})^2, \ldots,(\omega^{\frac{n}{d}})^{n-1}\Big)^T
    = \frac{1}{\sqrt{n}} \Big(1,-1,1,-1, \ldots, 1,-1 \Big)^T.
\end{eqnarray}

\begin{Prop} \label{Pro2} Let $n=2p$ and $k=2q$. If the map
\begin{equation}
   \tau'_{n,k}(X) =  \tau_{n,k}(X)- t v_1 v_1^\dagger \circ X ,
\end{equation}
is positive, then $t \leq n-k$.  
\end{Prop}
Proof: let $\mu = (1,0,1,0,\ldots,1,0)^T \in \mathbb{C}^n , $
and let $X = \mu \mu^\dagger$. Consider now the $p \times p$ submatrix $\mathbf{N}$ of $\tau'_{n,k}(X)$ consisting of even rows and columns, that is, $\mathbf{N}_{ij} = [\tau'_{n,k}(X)]_{2i,2j}$. One easily finds 
\begin{equation}\label{}
 \mathbf{N} = (2p-q)  \mathbf{I}_p -  \left(1+\frac{t}{n}\right) \mathbb{J}_p .
\end{equation}
The eigenvalue of $\mathbf{N}$ corresponding to the eigenvector $(1,1,\ldots,1)^T$ reads
$$  (2p-q) - p\left(1+\frac tn\right) = \frac 12 (2(p-q) - t) = \frac 12 ((n-k)-t) , $$
and hence $t \leq n-k$ to ensure positivity. \hfill $\Box$

\begin{Conj}
If  ${\rm gcd}(n,k)=2$, we conjecture that
\begin{equation}
   \tau'_{n,k}(X) =  \tau_{n,k}(X)- (n-k) v_1 v_1^\dagger \circ X ,
\end{equation}
is a positive optimal map. 
It is clear, that whenever $\tau'_{n,k}$ is positive it has to be optimal since there is no room for subtraction of another CP map. Our conjecture is strongly supported by the numerical analysis.

\end{Conj}
In particular, for ${\rm gcd}(4,2)=2$, it was proved \cite{PRA-2022}  that the map
\begin{equation}\label{}
  \tau_{4,2}(X) =  \begin{bmatrix}  x_{00} + x_{11}  + x_{22} & - x_{01} & - x_{02} & - x_{03} \\ 
  -x_{10} &  x_{11} + x_{22}  + x_{33} & - x_{12} & - x_{13} \\ -x_{20} & - x_{21} &  x_{22} + x_{33}  + x_{00} & - x_{23} \\ -x_{30} & -x_{31} & - x_{32} &  x_{33} + x_{00}  + x_{11}  \end{bmatrix} ,
\end{equation}
is not optimal. However, the {\it corrected}  map
\begin{equation}\label{}
  \tau'_{4,2}(X) =  \begin{bmatrix}   \frac 12 x_{00} + x_{11}  + x_{22} & - \frac 12  x_{01} & - \frac 32 x_{02} & - \frac 12 x_{03} \\
  - \frac 12 x_{10} & \frac 12 x_{11} + x_{22}  + x_{33} & - \frac 12 x_{12} & - \frac 32 x_{13} \\ - \frac 32 x_{20} & - \frac 12 x_{21} & \frac 12 x_{22} + x_{33}  + x_{00} & - \frac 12 x_{23} \\ - \frac 12 x_{30} & -\frac 32 x_{31} & - \frac 12 x_{32} & \frac 12 x_{33} + x_{00}  + x_{11}  \end{bmatrix} ,
\end{equation}
 is positive. Proposition \ref{Pro2} implies

\begin{Cor} The map $ \tau'_{4,2}$ is optimal. 
\end{Cor}

\section{Conclusions} 

In this paper, it is proved that the class of positive maps $\tau_{n,k}$ is optimal whenever $\mathrm{gcd}(n,k)=1$. This class of maps provides generalization of a seminal Choi map $\tau_{3,1}$ which was proved to be extremal and hence optimal. In particular, for any $n$, all maps $\tau_{n,1}$ are optimal and for odd $n$, all maps $\tau_{n,n-2}$ are optimal.
We have proved that maps $\tau_{n,k}$ do not have a spanning property (apart from the reduction map $\tau_{n,n-1}$). The optimality of  maps without the spanning property is rather exceptional. Besides Choi map $\tau_{3,1}$ and $\tau_{4,2}$ (analyzed recently in  \cite{PRA-2022}), we are aware of only one additional example constructed in \cite{Remik}.

If $\mathrm{gcd}(n,k) >1$, then in general $\tau_{n,k}$ is not optimal. In particular, for  $\mathrm{gcd}(n,k) =2$, we have provided a conjecture which says that the map $\tau_{n,k}$ can be optimized by subtracting a CP map being a Hadamard product with $(n-k) v_1 v_1^\dagger$. We have shown it for $(n,k)=(4,2)$ scenario. For $\mathrm{gcd}(n,k) > 2$, the situation is more complicated since there is two-dimensional kernel of the matrix $\mathbf{M}$ and there is more freedom to subtract CP maps.

It would be very interesting to analyze which optimal maps $\tau_{n,k}$ are also extremal. Here we propose the following 

\begin{Conj} $\tau_{n,k}$ is extremal if and only if ${\rm gcd}(n,k)=1$. 
\end{Conj}
We postpone this problem for the future research.

\section*{Acknowledgements}
We would like to thank the anonymous referee for his/her fruitful comments which helped us to improve the clarity of the paper. The work was supported by the Polish National Science Centre project No. 2018/30/A/ST2/00837.

\appendix
\section{Simple cases for Theorem \ref{TH-M}}
\label{simple}
\subsection{The map $\tau_{3,1}$}
First we consider one of the simple example $\tau_{3,1}$, for $n=3$ and $k=1$. Due to the Definition in Eq.~\eqref{!}, $\tau_{3,1}$ can be expressed as
\begin{equation}
\label{mat31}
  \tau_{3,1}(X) = \begin{bmatrix}
   2x_{00} + x_{11} & \cdot & \cdot \\ \cdot & 2x_{11}+x_{22} & \cdot \\ \cdot & \cdot & 2 x_{22}+x_{00} 
\end{bmatrix}-X.
\end{equation}
Now, according to Eqs.~\eqref{taualpha} and \eqref{u}, we take $X= \begin{bmatrix}
  1 & s & \cdot \\ s & s^2 & \cdot \\ \cdot & \cdot & \cdot 
\end{bmatrix}.$ Then
\begin{equation}
    \tau_{3,1}^\alpha(X) = \begin{bmatrix}
   2 +s^2 & \cdot & \cdot \\ \cdot & 2 s^2 & \cdot \\ \cdot & \cdot & 1 
\end{bmatrix}-\begin{bmatrix}
  1 & s & \cdot \\ s & s^2 & \cdot \\ \cdot & \cdot & \cdot 
\end{bmatrix}-\alpha\alpha^\dagger \circ \begin{bmatrix}
  1 & s & \cdot \\ s & s^2 & \cdot \\ \cdot & \cdot & \cdot 
\end{bmatrix} = \begin{bmatrix} A & \cdot \\ \cdot & B 
      \end{bmatrix} ,
\end{equation}
where $B \in M_{1}$ is evidently  positive semi-definite, and $A \in M_{2}$ is defined as follows
\begin{equation}
    A=\begin{bmatrix} 2+s^2 & \cdot \\ \cdot & 2 s^2 \end{bmatrix} - \begin{bmatrix} 1 & s \\ s & s^2 \end{bmatrix} -\alpha\alpha^\dagger \circ \begin{bmatrix}
  1 & s \\ s & s^2 \end{bmatrix}=\begin{bmatrix}
      1 & \cdot \\ \cdot & s \end{bmatrix} 
         \left( 2 \mathbb{I}_2 + \begin{bmatrix} s^2 & \cdot \\ \cdot & \cdot \end{bmatrix} - \mathbb{J}_2 -\widetilde{\alpha} \widetilde{\alpha}^\dagger \right)
      \begin{bmatrix}
      1 & \cdot \\ \cdot & s \end{bmatrix}.
\end{equation}
Now, the matrix in the middle has to be positive semi-definite for all values of $s$, hence also in the limit of  $s \to 0$: $2 \mathbb{I}_2 - \mathbb{J}_2 -\widetilde{\alpha} \widetilde{\alpha}^\dagger \ge 0$. The matrix 
$2 \mathbb{I}_2 - \mathbb{J}_2$ is a projector of the orthogonal complement of one-dimensional subspace spanned by $[1,1]$, hence $2\mathbb{I}_2 - \mathbb{J}_2 -\widetilde{\alpha} \widetilde{\alpha}^\dagger$ is positive-semidefinite iff $[1,1] \widetilde \alpha = 0 \Leftrightarrow [1,1,0] \alpha = 0$.

Proceeding analogously taking $X$ to be projector on vectors $[0,1,s]$ and $[s,0,1]$ we obtain a system of equations:
\begin{equation}
    \begin{bmatrix}
        1 & 1& \cdot \\   \cdot & 1 & 1 \\   1 &  \cdot & 1
    \end{bmatrix} \alpha=0 \implies \alpha=0,
\end{equation}
due to non-singularity of the system matrix. Hence no map of the type \eqref{lambdax} can be subtracted from 
 the original map $\tau_{3,1}$. It proves that the map $\tau_{3,1}$ is optimal.

\subsection{The map $\tau_{4,3}$}
Now we consider the map $\tau_{4,3}$ for $n=4$ and $k=3$ (the reduction map). From the definition, $\tau_{4,3}$ can be expressed as
\begin{equation}
    \tau_{4,3}(X)=\begin{bmatrix}
   x_{00}+x_{11}+x_{22}+x_{33} & \cdot & \cdot & \cdot \\ 
   \cdot & x_{11}+x_{22}+x_{33}+x_{00} & \cdot & \cdot \\ 
   \cdot & \cdot & x_{22}+x_{33}+ x_{00}+x_{11} & \cdot \\  
    \cdot & \cdot & \cdot & x_{33}+ x_{00}+x_{11}+x_{22}
    \end{bmatrix}-X.
\end{equation}
Just like the above subsection, according to Eqs.~\eqref{taualpha} and \eqref{u}, we take $X= \begin{bmatrix}
  1 & \cdot & \cdot & \cdot \\ \cdot & \cdot & \cdot & \cdot \\ \cdot & \cdot & \cdot & \cdot \\  \cdot & \cdot & \cdot & \cdot 
\end{bmatrix}=e_{00}.$
Then
\begin{equation}
    \tau_{4,3}^\alpha(X) = \begin{bmatrix}
   1 & \cdot & \cdot & \cdot \\ \cdot & 1 & \cdot & \cdot \\ \cdot & \cdot & 1 & \cdot \\  \cdot & \cdot & \cdot & 1 
\end{bmatrix}-\begin{bmatrix}
 1 & \cdot & \cdot & \cdot \\ \cdot & \cdot & \cdot & \cdot \\ \cdot & \cdot & \cdot & \cdot \\  \cdot & \cdot & \cdot & \cdot 
\end{bmatrix}-\alpha\alpha^\dagger \circ \begin{bmatrix}
  1 & \cdot & \cdot & \cdot \\ \cdot & \cdot & \cdot & \cdot \\ \cdot & \cdot & \cdot & \cdot \\  \cdot & \cdot & \cdot & \cdot 
\end{bmatrix} = \begin{bmatrix} A & \cdot \\ \cdot & B 
      \end{bmatrix} ,
\end{equation}
where $B \in M_{3}$ is evidently  positive semi-definite, and $A \in M_{1}$ can be written as 
$A=-\alpha\alpha^\dagger \circ [1]$. Therefore $-\alpha\alpha^\dagger \geq 0 \Leftrightarrow \Tilde{\alpha}=0 \Leftrightarrow [1,0,0,0] \alpha=0$. Proceeding analogously taking $X$ to be $e_{11}$, $e_{22}$, and $e_{33}$, we obtain a system of
equations:
\begin{equation}
    \begin{bmatrix}
  1 & \cdot & \cdot & \cdot \\ \cdot & 1 & \cdot & \cdot \\ \cdot & \cdot & 1 & \cdot \\  \cdot & \cdot & \cdot & 1 
\end{bmatrix} \alpha=0 \implies \alpha=0,
\end{equation}
due to non-singularity of the system matrix. Hence no map of the type \eqref{lambdax} can be subtracted from 
 the original map $\tau_{4,3}$. It proves that the map $\tau_{4,3}$ is optimal.

\subsection{The map $\tau_{5,3}$}
Now we consider the map $\tau_{5,3}$ for $n=5$ and $k=3$. From the definition, $\tau_{5,3}$ can be expressed as
\begin{eqnarray}
    \tau_{5,3}(X) &=& \begin{bmatrix}
  2 x_{00}+\sum_{i=1}^{3} x_{ii}& \cdot & \cdot & \cdot & \cdot \\ 
   \cdot & 2 x_{11}+\sum_{i=2}^{4} x_{ii} & \cdot & \cdot & \cdot \\ 
   \cdot & \cdot & 2 x_{22}+\sum_{i=3}^{4} x_{ii}+x_{00} & \cdot & \cdot \\  
    \cdot & \cdot & \cdot & 2 x_{33}+x_{44}+ \sum_{i=0}^{1} x_{ii} & \cdot \\
     \cdot & \cdot & \cdot & \cdot & 2 x_{44}+ \sum_{i=0}^{2} x_{ii}
    \end{bmatrix} \nonumber\\
    &-& X. \nonumber\\
\end{eqnarray}
according to Eqs.~\eqref{taualpha} and \eqref{u}, we take $X= \begin{bmatrix}
  1 & s & \cdot &  \cdot & \cdot \\ 
  s & s^2 & \cdot & \cdot & \cdot \\ 
  \cdot & \cdot & \cdot & \cdot & \cdot \\ 
  \cdot & \cdot & \cdot & \cdot & \cdot \\
  \cdot & \cdot & \cdot & \cdot & \cdot 
\end{bmatrix}.$
Then
\begin{equation}
    \tau_{5,3}^\alpha(X) = \begin{bmatrix}
  2+s^2 & \cdot & \cdot &  \cdot & \cdot \\ 
  \cdot & 2 s^2 & \cdot & \cdot & \cdot \\ 
  \cdot & \cdot & 1 & \cdot & \cdot \\ 
  \cdot & \cdot & \cdot & 1+s^2 & \cdot \\
  \cdot & \cdot & \cdot & \cdot & 1 +s^2
\end{bmatrix}-\begin{bmatrix}
 1 & s & \cdot &  \cdot & \cdot \\ 
  s & s^2 & \cdot & \cdot & \cdot \\ 
  \cdot & \cdot & \cdot & \cdot & \cdot \\ 
  \cdot & \cdot & \cdot & \cdot & \cdot \\
  \cdot & \cdot & \cdot & \cdot & \cdot 
\end{bmatrix}-\alpha\alpha^\dagger \circ \begin{bmatrix}
  1 & s & \cdot &  \cdot & \cdot \\ 
  s & s^2 & \cdot & \cdot & \cdot \\ 
  \cdot & \cdot & \cdot & \cdot & \cdot \\ 
  \cdot & \cdot & \cdot & \cdot & \cdot \\
  \cdot & \cdot & \cdot & \cdot & \cdot  
\end{bmatrix} = \begin{bmatrix} A & \cdot \\ \cdot & B 
      \end{bmatrix} ,
\end{equation}
where $B \in M_{3}$ is evidently positive semi-definite, and $A \in M_{2}$ is defined as follows
\begin{equation}
    A=\begin{bmatrix} 2+s^2 & \cdot \\ \cdot & 2 s^2 \end{bmatrix} - \begin{bmatrix} 1 & s \\ s & s^2 \end{bmatrix} -\alpha\alpha^\dagger \circ \begin{bmatrix}
  1 & s \\ s & s^2 \end{bmatrix}=\begin{bmatrix}
      1 & \cdot \\ \cdot & s \end{bmatrix} 
         \left( 2 \mathbb{I}_2 + \begin{bmatrix} s^2 & \cdot \\ \cdot & \cdot \end{bmatrix} - \mathbb{J}_2 -\widetilde{\alpha} \widetilde{\alpha}^\dagger \right)
      \begin{bmatrix}
      1 & \cdot \\ \cdot & s \end{bmatrix}.
\end{equation}
Now, the matrix in the middle has to be positive semi-definite for all values of $s$, hence also in the limit of  $s \to 0$: $2 \mathbb{I}_2 - \mathbb{J}_2 -\widetilde{\alpha} \widetilde{\alpha}^\dagger \ge 0$. The matrix 
$2 \mathbb{I}_2 - \mathbb{J}_2$ is a projector of the orthogonal complement of one-dimensional subspace spanned by $[1,1]$, hence $2\mathbb{I}_2 - \mathbb{J}_2 -\widetilde{\alpha} \widetilde{\alpha}^\dagger$ is positive-semidefinite iff $[1,1] \widetilde \alpha = 0 \Leftrightarrow [1,1,0,0,0] \alpha = 0$.

Proceeding analogously taking $X$ to be projector on vectors $[0,1,s,0,0]$, $[0,0,1,s,0]$, $[0,0,0,1,s]$ and $[s,0,0,0,1]$ we obtain a system of equations:
\begin{equation}
    \begin{bmatrix}
       1 & 1 & \cdot &  \cdot & \cdot \\ 
  \cdot  & 1 & 1 & \cdot & \cdot \\ 
  \cdot & \cdot & 1 & 1 & \cdot \\ 
  \cdot & \cdot & \cdot  & 1 & 1 \\
  1 & \cdot & \cdot & \cdot & 1  
    \end{bmatrix} \alpha=0 \implies \alpha=0,
\end{equation}
due to non-singularity of the system matrix. Hence no map of the type \eqref{lambdax} can be subtracted from 
 the original map $\tau_{5,3}$. It proves that the map $\tau_{5,3}$ is optimal.

\end{document}